\documentstyle[11pt]{article}
\newcommand\bc{\begin{center}}
\newcommand\ec{\end{center}}
\title{The Limits on Cosmological Anisotropies and \\ Inhomogeneities from
	COBE Data}
\author{}
\date{}
\begin{document}
\maketitle
\bc
William R. Stoeger$^*$, Marcelo E. Araujo$^{@*}$ and Tim Gebbie$^+$
\ec
\medskip
$^*$ Vatican Observatory Research Group, Steward Observatory, University of
Arizona, Tucson, Arizona  85721, U. S. A. \\

\noindent
$^@$ Departmento de Matem\'{a}tica, Universidade de Brasilia - UnB, 
70.910-900 Brasilia - D. F., Brazil  \\

\noindent
$^+$ Department of Mathematics and Applied Mathematics, University of 
Cape Town, Rondebosch 7700, Cape Town, South Africa

\begin{abstract}
Assuming that the cosmological principle holds, Maartens, Ellis and Stoeger
(MES) recently constructed a detailed scheme
linking anisotropies in the
cosmic background radiation (CMB) with anisotropies and inhomogeneities in the
large scale structure of the universe and showed how to place limits on those
anisotropies and inhomogeneities simply by using CMB quadrupole and octupole 
limits. First we indicate and discuss the connection between the covariant 
multipole moments of the temperature anisotropy used in the MES scheme and 
the quadrupole
and octupole results from COBE. Then we introduce those results into the
MES limit equations to obtain definite quantitative limits on the complete
set of cosmological measures of anisotropy and inhomogeneity. We find that 
all the anisotropy measures are less than $10^{-4}$ in the case of those
not affected by the expansion rate $H$, and less than $10^{-6} \; Mpc^{-1}$ in 
the case of those which are. These results quantitatively demonstrate that
the observable universe is indeed close to Friedmann-Lema\^{i}tre-Robertson-
Walker (FLRW) on the largest scales, and can be adequately modelled by an
almost-FLRW model -- that is, the anisotropies and inhomogeneities
characterizing the observable universe on the largest scales are not too
large to be considered perturbations to FLRW.
\end{abstract}

\noindent
{\it Subject Headings:} cosmology: theory -- cosmology: observations --
cosmic microwave background radiation -- relativity \\

\section{Introduction}

People have long considered that the almost isotropy of the cosmic microwave
background radiation (CMB) indicates that the universe on very large scales is
isotropic and homogeneous, or nearly so. The first rigorous result supporting
this conjecture was the theorem proved by Ehlers, Geren and Sachs (1968) (EGS):
``If
a family of freely falling observers measure self-gravitating background
radiation to be everywhere exactly isotropic, then the universe is exactly
Friedmann-Lema\^{i}tre-Robertson-Walker (FLRW).'' Later, Grishchuk and
Zel'dovich (1978), analyzing the FLRW perturbations, argued that, if CBR
anisotropies are very small, all anisotropies and inhomogeneities on scales
larger than the horizon should also be very small. Much more recently Stoeger
et al. (1995) (SME)  proved a significant generalization of this theorem:
``If the
Einstein-Liouville equations are satisfied in an expanding universe, where 
there is present pressure-free matter with 4-velocity vector field $u^a$
($u_au^a = -1$) such that (freely propagating) background radiation is
everywhere almost-isotropic relative to $u^a$, then spacetime is almost
FLRW.'' These results obviously provide a fundamental link between observational
and theoretical cosmology -- one which promises to reveal key aspects of the
very
large scale
structure of the universe without relying on the often uncertain observational
measurements of local and intermediate scale structures. It should be noted
that both the EGS and the SME results depend on assuming that the cosmological
principle holds -- here expressed in terms of the almost-isotropy of the
background radiation relative to {\it every} fundamental observer in the
space-time. Goodman (1995) and Maartens et al. (1995a), however, have recently
pointed
out the important fact that
the cosmological principle itself is partially testable via the
Sunyaev-Zel'dovich effect. \\

Employing this connection between CBR anisotropies and the full range of
cosmological isotropies and inhomogeneities, Maartens et al. (1995a,b) (MES)
have
developed a
detailed
scheme demonstrating how limits on the temperature anisotropies of the
CMB imply rigorous limits on the anisotropy and inhomogeneity of the
universe. In this paper we show how CMB anisotropy data is inserted into
the theoretically derived limits of this scheme to constrain strongly the
vorticity, shear, spatial gradients, Weyl-tensor components, and other
measures of deviation from FLRW for the observable universe. We then introduce
the limits
COBE places on the dipole, quadrupole and octupole of the CBR to determine
those limits quantitatively. \\

In the next section we shall briefly give the key MES equations which 
constitute the CMB limits on the anisotropy and inhomogeneity of the universe.
Then, in the third section, we shall discuss the relationship of the
usual CMB multipole results with the temperature anisotropy multipoles in the
MES equation. Finally, in section 4, we shall present the COBE quadrupole
and octupole observational results, and use these to arrive at limits on
all the cosmological anisotropy and inhomogeneity measures mentioned above. \\

Maartens, Ellis and Stoeger (1996) recently provided such limits by introducing
order-of-magnitude COBE results of $\epsilon_2 \approx
\epsilon_3 \approx 10^{-5}$, but they
did not discuss the relationship between temperature anisotropy multipoles
used in their scheme and those in terms of which the COBE results are given.
Here we fill that gap, and then use the up-to-date COBE results for the rms
quadrupole and octupole to obtain improved limits.

\section{The MES Equations}

MES have written the covariant temperature isotropy
multipoles as $\tau_{a_1 ...a_L}$, where $L$ give the multipole number.
Thus, for instance, $\tau_a$, $\tau_{ab}$, $\tau_{abc}$ are, respectively,
the components of the dipole, quadrupole, and octopole of the CMB 
temperature anisotropy. This form of the harmonic decomposition, which is
given in detail in Maartens et al. (1995a), is equivalent to that formulated
in terms of spherical harmonics (Ellis, Matravers and Treciokas 1983; 
Ellis Treciokas and Matravers 1983). \\

MES have assumed that there are observed bounds on these temperature anisotropy
multipoles, so that there exist O[1] constants $\epsilon _L$ such that
$$\mid \tau_{a_1...a_L} \mid < \epsilon_L.  \eqno(1) $$
The absolute-value brackets have been {\it defined} to be the squareroot of
the sum of the squares of the components of a given vector or tensor
(Maartens et al. 1995a, p. 1526). Thus, $\epsilon_1$ gives the limits on 
the dipole components, $\epsilon_2$ on the quadrupole components, 
$\epsilon_3$ on the octupole moments, and so on. \\

Then, using the strong observational assumptions on the spatial gradients and
time-derivatives of the temperature harmonics they adopted in Maartens et al. 
(1995b), they have the observational limit equations on various kinematic, 
dynamic and geometric indicators of anisotropy and inhomogeneity 
(Maartens et al. 1995a,b):

$$\frac{\mid \hat{\bigtriangledown}_a \mu \mid}{\mu} = 4\frac{\mid \hat{
\bigtriangledown}_a T \mid}{T} < H (12 \epsilon_1 + \frac{24}{5} \epsilon_2 ),
\eqno (2) $$

$$\frac{\mid \sigma_{ab} \mid}{\Theta} < \frac{5}{3}\epsilon_1 + 3 \epsilon_2
    + \frac{3}{7} \epsilon_3, \eqno(3) $$

$$\frac{\mid \omega_{ab} \mid}{\Theta} < \frac{10}{3}\epsilon_1 +
    \frac{2}{15} \epsilon_2, \eqno(4) $$

$$\frac{\mid \hat{\bigtriangledown}_a \rho \mid}{\Theta} < \frac{9}{2}H
 \epsilon_2 + (\frac{H}{\Omega_M})[60 \epsilon_1 + 134 \epsilon_2 +
 6\epsilon_3] + (\frac{\Omega_R}{\Omega_M})H[16 \epsilon_1 + \frac{61}{3}
 \epsilon_2], \eqno(5) $$

$$\frac{\mid \hat{\bigtriangledown}_a \Theta \mid}{\Theta}
< H(\frac{205}{3}
\epsilon_1 + 8 \epsilon_2) + 4(2 \Omega_R + \Omega_M) H \epsilon_1, \eqno(6)$$

$$\frac{\mid E_{ab} \mid}{\Theta} < H(\frac{55}{3} \epsilon_1 + \frac{103}{3}
 \epsilon_2 + \frac{23}{7} \epsilon_3) + \frac{4}{45}(11 \Omega_R +
 15 \Omega_M)H\epsilon_2, \eqno(7) $$

$$\frac{\mid H_{ab} \mid}{\Theta} < H(\frac{16}{3} \epsilon_1 + \frac{52}{15}
\epsilon_2 + \frac{1}{21} \epsilon_3). \eqno(8) $$

\noindent
Here $\mu$ and $\rho$ are the radiation and matter densities, respectively,
and $\Omega_R$ and $\Omega_M$ are the ratios of the radiation energy density
and the matter densities, respectively, to the critical density of the 
universe. $H$, of course, the Hubble parameter, and $T$ is the CMB temperature.
$\sigma_{ab}$ is the shear of the congruence of timelike geodesics in the
universe, $\omega_{ab}$ its vorticity,
$\Theta = u^a_{;a}$ its expansion scalar, and $E_{ab}$ and $H_{ab}$ the
electric and magnetic components, respectively, of the Weyl tensor. These
quantities, along with the spatial gradients of $\rho$, $\mu$ and $\Theta$
describe the anisotropy and inhomogeneity of the space-time. The $\hat
{\bigtriangledown}_a$ operator expresses the covariant spatial gradient in
the spacetime. It is defined by

$$\hat{\bigtriangledown}_c Q_{a...b} \equiv h_c^d h_a^e h_b^f \bigtriangledown
_d Q_{e...f},$$
where $h_{ab} = g_{ab} + u_au_b$ is the projection tensor in the rest spaces of
the geodesically moving observers and $\bigtriangledown_a$ is the covariant
derivative define by the metric $g_{ab}$. Finally, it should be noted that
equations (2) - (8) hold independently of the statistics of the underlying
matter-density fluctuations -- they do not assume that the fluctuations are
Gaussian. \\

If we can determine $\epsilon_1$, $\epsilon_2$ and $\epsilon_3$ from CMB
measurements, and have observational estimates of $H$, $\Omega_R$ and
$\Omega_M$ from other observations, then we can determine limits on all these
anisotropy and inhomogeneity indicators. We shall proceed to do this in 
Section 4. But first, in the next section, we need to discuss the relationship
between the $\mid\tau_{a_1...a_L} \mid$ given in the equations and the multipole
moment results determined by the COBE Differential Microwave Radiometers
(DRM) and other CMB anisotropy detectors. 

\section{CMB Multipole Anisotropy Data}

We have mentioned above that the harmonic decomposition represented by the
$\mid\tau_{a_1...a_L}\mid$ is equivalent to that in terms of spherical
harmonics.
But the question then is whether or not the multipole results
recovered from, say, COBE DRM data can be simply substituted into our limit
equations. Are the multipole results presented in the COBE papers measurements
of the $\mid\tau_{a_1...a_L}\mid$ in our limit equations? The answer to this
question
is `Yes,' as long as we use the real rms dipole, quadrupole,
octupole moments they obtain, and not those associated with obtaining the
power spectrum, which are often the focus of their reported results, determine
the numerical factors relating these moments, defined in terms of Legendre
polynomials, to the $\mid \tau_{a_1 . . . a_L} \mid$ (see section 5 below), and
as long as we realize that they are usually given as the squareroot of the sum
of the squares of the $(2L + 1)$
components of the L-pole, the rms L-pole  -- not as values of each separate
component
of the multipole in question. Furthermore, in the COBE data the multipoles
are quoted for $\delta T$, instead of $\frac{\delta T}{T}$, for which our
$\mid\tau_{a_1...a_L}\mid$ are the multipoles. The COBE rms values published
all have
units of $\mu K$, therefore. To translate these values into what we need we
must thus divide them by $T$, the average CMB background temperature over the
sky. \\

As a relevant example, Bennett el al. (1994) 
give
results for the square of the rms quadrupole amplitude $Q^2_{rms}$,

$$Q^2_{rms} = \frac{4}{15}(\frac{3}{4}Q_1^2 + Q_2^2 + Q_3^2 + Q_4^2 + Q_5^2),
  \eqno(9)$$

\noindent
where the five quadrupole components $Q_i$ are given in terms of Galactic
coordinates $l$ and $b$ (Galactic longitude and latitude respectively) by the
expansion

\setcounter{equation}{9}
\begin{eqnarray}
Q(l, b) &  = &  Q_1(3 \sin^2 b - 1)/2 + Q_2 \sin 2b \cos l + Q_3 \sin 2b \sin l
+ \nonumber \\
   & & Q_4 \cos^2 b \cos 2l + Q_5 \cos^2 b \sin 2l. 
\end{eqnarray}

\noindent
The `strange' coefficients in equation (9) are due to the fact that the
basis vectors are orthogonal but not orthonormal (see Partridge 1995);
$Q^2_{rms}$ does
{\it not} contain a factor of $(2L + 1)^{-1} = 1/5$. 
>From four years of
data the COBE workers
give us a best fit value of (Smoot, private communication; Kogut et al., 1996) 
$$Q_{rms} = 10.7 \pm 7 \mu K , \eqno(11)$$
with a 95\% confidence limit. We can modify this result (see below) for
use in equations (2) to (8). This quantity is to be carefully
distinguished in the COBE results from $Q_{rms-PS}$ which is often referred
to and which is {\it not} the true best fit value of the quadrupole, but rather
the value of the quadrupole derived from a power-spectrum fit (when a power law
is assumed) of the other
higher-order multipole moments [see Bennett et al. 1994; Smoot et al. 1992] \\

The dipole moment can be neglected, according to the fairly well justified
assumption that it is all
due to our peculiar motion with respect the rest frame of the microwave
background. In fact this is what is done in the COBE anisotropy analysis
(Bennett et al. 1994).
However, we should be aware that there could in principle be a small non-
Doppler contribution to the CMB dipole (see Maartens et al. 1995a and Maartens
et al. 1996 concerning this). Thus, in
our
calculations below we shall set
$$\epsilon_1 = 0. \eqno(12)$$ \\

The COBE workers have not yet published the rms
octupole from their data, but G. Smoot (private communication) has kindly
informed us that the rms octupole results from COBE are: 
$$ O_{rms} = 16 \pm 8 \mu K. \eqno(13)$$ \\ 
This is consistent with the results given by Wright et al. (1994) in their
Figure 1. \\

There are several important observational and data-reduction issues, and
one theoretical issue, which we should briefly mention here, to provide
the background against which we can understand and appreciate these COBE
multipole results. \\

The first is that the COBE DMR experiment does not directly measure the
dipole, quadrupole, octupole moments of the temperature anisotropy, but
rather the two-point correlation function of the temperature anisotropy (Smoot
et al. 1992, Padmanabham 1993, Partridge 1995)
$$C(\alpha) = <S({\bf n})S({\bf m})> = \sum_{\ell = 1}^\infty \bigtriangleup
  T_{\ell}^2 W(\ell)^2 P_{\ell}[\cos \alpha], \eqno(14)$$
where $\alpha$ is the angle between the two points, ${\bf n}$ and ${\bf m}$
are unit vectors
denoting the two different directions, so that $\cos \alpha
= {\bf n} \cdot {\bf m}$, and the averaging brackets signify the average over
all pairs of points on the sky with separation angle $\alpha$. Furthermore,
$S({\bf n}) \equiv \delta T
({\bf n})$, the temperature anisotropy in a given direction ${\bf n}$ on the
plane of the sky. The $\bigtriangleup T_{\ell}^2$ (with $\ell \equiv L$) are 
the squares of the rotationally invariant rms multipole moments 
(thus, $\bigtriangleup T_2^2 = Q^2_{rms}$ -- see Bennett et al. 1994, for 
instance) $$\bigtriangleup T_{\ell}^2 = \frac{1}{4 \pi} \sum_m 
\mid a_{\ell m} \mid^2, \eqno(15)$$ where the $a_{\ell m}$ are the 
coefficients of the expansion of the temperature anisotropy in spherical 
harmonics, i. e.
$$S({\bf n}) \equiv S(\theta, \phi) = \sum_{\ell,m} a_{\ell m} 
Y_{\ell m}(\theta, \phi), \eqno(16) $$
$\theta$ and $\phi$ being, of course, the angular coordinates of the 
direction at which the temperature anisotropy is being measured. In 
equation (14), finally, $P_{\ell}[\cos \alpha]$ is just the Legendre 
polynomial of degree $\ell$ given as a function of $\cos \alpha$, and 
$W(\ell)$ is the window function, which describes the smoothing 
properties of the instrument's beam. For detectors measuring large-angle 
CMB anisotropies it essentially weights the multipole moments in such 
a way that the higher multipoles are smoothed over -- that is, the 
instument is insensitive to anisotropies on angular scales less than 
a certain $\ell$-pole, and $W(\ell)$ describes that insensitivity and 
resulting transfer of power in the measurement from higher multipoles 
to lower ones. Thus, when the actual quadrupole or octupole is determined 
from the data, $\bigtriangleup T_\ell$ must be deconvolved from $W(\ell)$. 
When the rms dipole, quadrupole, octupole results are given by the COBE 
researchers, this deconvolution has already been performed. This is one 
of the procedures which must be effected to give us the real rms 
multipole moments. \\ 

The second major observational and data-reduction problem is that in the
COBE measurements there is
a great deal of contamination by experimental systematic errors, including
Galactic emission (see Bennett et al. 1994  and Smoot et al. 1992, and
references therein) And the lower
multipole moments are the most susceptible to these distortions. Furthermore,
when the data is processed, the entire region containing the Galaxy is removed
from the data set. This `Galactic cut' destroys the orthogonality of the
spherical harmonics and leads to further aliasing of higher order multipole
power onto the lower multipoles (dipole, quadrupole, octupole, etc.).
Corrections for this are estimated on the basis of Monte Carlo simulations 
(Bennett et al. 1994)
and included in the published values for the rms multipoles. There are 
a number of other complex issues which it has been necessary to resolve in
arriving at these values ( see Bennett et al. 1994, Smoot et al. 1992, and
Wright et al. 1994 for further discussion). \\

Finally, there is the theoretical-observational issue of `cosmic variance.' 
Actually, cosmic variance does not affect what we are concerned with
here, as we shall see. But it is important to realize why it does not. It may
explain why the multipoles we measure have the values they have relative to
theoretical models of the perturbation spectrum, but it is
does not lead to observational errors, which would have to be corrected for. It
{\it does} affect the comparison of the measured multipole power
spectrum with the theoretical power spectrum predicted from, say, inflationary
models (Abbott \& Wise 1984). If the primordial perturbation spectrum
originated due to 
fluctuations in the inflaton field, as we think it did, then the values of
the temperature anisotropy  multipoles they induce will be random variables
with a certain distribution, probably Gaussian with zero mean (Liddle \& Lyth
1993), and thus
with a
certain variance. Our observable universe is only one realization of that
ensemble of universes represented by the probability distribution. Therefore, 
the value for each multipole we obtain from our observations will give us just
one point of the distribution, which, in general, will not reflect the
ensemble averaged value. It will deviate from it by a certain amount, which
can be theoretically estimated by the variance (Liddle \& Lyth 1993). This
variance goes as $2/(2 \ell + 1)$ and so will be more significant for the 
lower multipoles (Abbott \& Wise 1984; Smoot et al. 1992).  Our concern 
here, however, is not to compare the observed power spectrum of CMB 
anisotropies with the theoretical spectrum of density perturbations
generated by an inflationary scenario. It is merely to use the best values
of the CMB multipole moments we have available -- however they are generated
and whatever their spectrum -- to set definite limits on the large scale
anisotropy and inhomogeneity of the observable universe itself. 
Thus, cosmic variance falls outside those issues which we need to 
consider in arriving at those limits. \\

\section{Calculating the Anisotropy and Inhomogeneity Limits}

We are now ready to use the values for the rms dipole, quadrupole and octupole
the COBE team have so far obtained to determine the anisotropy and
inhomogeneity of the universe on very large scales. As indicated above, we
set the dipole equal to zero -- equation (12). Thus, we shall set
$\epsilon_1 = 0$ in our equations (2) to (8). In order to transform 
equations (11) and (13) into values of $\mid \tau_{ab} \mid$ and $\mid
\tau_{abc} \mid$, respectively,  we need to divide the results in 
equations (11) and (13) by $T = 2.73 K$, since our multipoles are 
for $\delta T/T$.  We do not need to divide them by $(2 \ell + 1)^{1/2}$, 
since our multipole quantities are given as the absolute values, which 
we {\it defined} as the squareroot of the sum of the squares of the 
components (see above). Finally, we also need to relate the 
$\bigtriangleup T_{\ell}^2$, the coefficients in the usual Legendre 
polynomial expansion, equation (14) to the $\mid \tau_{a_1. . .a_{\ell}}\mid$. 
The numerical relationship is (Gebbie \& Ellis 1998)

$$\left < {\tau_{A_\ell} \tau^{A_\ell}} \right > = 
(2 \ell + 1){{(2 \ell )!\over 2^{\ell} ({\ell}!)^2}} 
 \Delta T_{\ell}^2. \eqno (17)$$

This gives

$$ \mid \tau_{ab} \mid^2 = 7.5 Q_{rms}^2, \eqno (18) $$
and
$$ \mid \tau_{abc}\mid^2 = 17.5 O_{rms}^2, \eqno (19)$$

We obtain, therefore,

$$ \langle \epsilon_2 \rangle = 1.1 (\pm 0.8) 
\times 10^{-5}, \eqno(20)$$
and
$$ \langle \epsilon_3 \rangle = 2.5 (\pm 1.3) 
\times 10^{-5}. \eqno(21)$$ \\

In examining equations (2) to (8), we notice that they are not
expressed in terms of average values of the components. Let us now consider
them to be equations for the rms averages of the indicators they
represent, which we can do, since our $\epsilon$'s are all positive. Now
writing the Hubble parameter as $H = 100 h \; km/sec/Mpc $, $0.4 < h < 1.0$,
and neglecting terms containing the factor $\Omega_R$, since this is 
presently so small -- $\Omega_R = 4.11 h^{-2} \times 10^{-5}$  
(Roos, 1994) -- we obtain:


$$\langle \frac{\vert \hat{\bigtriangledown}_a \mu \vert}{\mu}\rangle  <  
 1.8h \times 10^{-8} \; Mpc^{-1}, \eqno(22)$$

$$\langle \frac{\vert \sigma_{ab} \vert}{\Theta}\rangle  < 
 4.4 \times 10^{-5}, \eqno(23)$$

$$\langle \frac{\vert \omega_{ab} \vert}{\Theta}\rangle  < 
 1.5 \times 10^{-6}, \eqno(24) $$

$$\langle \frac{\vert \hat{\bigtriangledown}_a \rho \vert}{\Theta}\rangle  <   
 (0.02 + 0.54 \Omega_M^{-1} ) h \times 10^{-6} \; Mpc^{-1}, \eqno(25)$$

$$\langle \frac{\vert \hat{\bigtriangledown}_a \Theta \vert}{\Theta}\rangle  <  
   3.0 h \times 10^{-8} \; Mpc^{-1}, \eqno(26)$$

$$ \langle \frac{\vert E_{ab} \vert}{\Theta}\rangle < 
  (1.5 + 48 \Omega_M) h \times 10^{-7} \; Mpc^{-1}, \eqno(27)$$

$$\langle \frac{\vert H_{ab} \vert}{\Theta}\rangle  <
   1.6 h \times 10^{-8} \; Mpc^{-1},  \eqno(28)$$

\noindent
In equations with $H$ we have put a factor of $c^{-1}$ back into the equations
to give the correct order of magnitude and the correct dimensions. This
factor is hidden in the normalization of the 4-velocity $u^a$, which figures 
implicitly in the equations. \\

These are the limits we desire on the anisotropies and the inhomogeneities 
of the universe. In the past other groups have put limits on the shear and 
the vorticity using limits on CBR isotropies (Bajtlik et al. 1986; Martinez-
Gonzalez \& Sanz 1995; Barrow et al. 1985). But in doing so they assumed 
exact spatial homogeneity, which yields a limit on shear which is too
strong (Maartens et al. 1996). Our limiting scheme, as we have mentioned, 
does not assume exact homogeneity, or even that the inhomogeneities and 
anisotropies are small. The result that they are small is due to the fact 
that the CBR anisotropies are small. The primary assumption we have made is 
a weak form of the Copernican principle -- that all fundamental observers 
in the relevant space-time domain measure at most the same level of CBR 
anisotropy. This, as we have pointed out, is at least partially testable -- 
and fully falsifiable. A single observation demonstrating that CBR 
anisotropies relative to some other cluster of galaxies is large compared 
to those we observe would banish that assumption. Finally, our approach 
provides limits on the full range of possible isotropies and 
inhomogeneities, including spatial gradients of the radiation and matter 
densities, and of the expansion parameter, and the electric and magnetic 
components of the Weyl tensor. It is worth pointing out that the Weyl 
tensor components measure those parts of the gravitational curvature which 
are not determined locally by the mass-energy distribution(that is, via 
the Einstein field equations). Instead, they are determined by the Bianchi 
identities, and so, in a sense, are due to the mass-energy distribution 
at other, more distant points (Hawking \& Ellis 1973; Maartens et
al. 1996). $H_{ab}$, the magnetic Weyl tensor, measures the amount of 
gravitational radiation in the space-time. It vanishes in the case in which 
exact spatial homogeneity and isotropy are assumed. $E_{ab}$, the electric 
Weyl tensor, which also vanishes in that case, measures the tidal, 
shear-inducing force of the global gravitational field. It thus manifests 
its presence by the shear it introduces in timelike and null congruences 
of geodesics -- worldlines of particles. In fact, the measurement of null 
shear in bundles of light rays would be the clearest signature of the 
presence of non-zero $E_{ab}$. (see Hawking \& Ellis 1973) \\

We see clearly from our results, equations (19) to (25), that in every case
the anisotropy and inhomogeneity measures for the universe on the largest 
scales are very small, despite the significant inhomogeneities which have 
been detected on intermediate scales. This provides very strong justification 
for considering the observable universe to be almost-FLRW on large 
scales -- that is, the deviations from FLRW are small enough to be 
considered perturbations. \\

Finally, we might wonder what the relevance of these limits is to ``the
averaging problem'' (see Zotov and Stoeger 1995 and references therein) in
cosmology. Do these results demonstrate that the universe is almost-FLRW
whatever the resolution of those issues happen to be? Unfortunately,
that is not the case. The analysis we have implemented here {\it assumes}
that the effective theory of gravity on cosmological scales is general
relativity. The averaging problem calls that assumption into question. \\

We are very grateful to George Smoot for providing recently derived COBE
values for the rms quadrupole and octupole, as well as for clarifying 
the principal source of error and the conventions in the COBE results. Special
thanks also to Roy Maartens and George Ellis for encouragement and helpful
comments. Marcelo Araujo thanks the Vatican Observatory Research Group
for support and hospitality during the course of this work.\\

\newpage

\noindent
{\bf REFERENCES} \\

\noindent
Abbott, L. F. \& Wise, M. B. 1984, ApJ, 282, L47. \\

\noindent
Bajtlik, S., Juszkiewicz, R., Pr\'{o}szy\'{n}dski, M., \& Amsterdamski, P.
1986, ApJ, 300, 463. \\

\noindent
Barrow, John D., Juszkiewicz, R., \& Sonoda, D. H. 1985, MNRAS, 213, 917. \\

\noindent
Bennett, C. L. et al. 1994, ApJ, 436, 423. \\

\noindent
Ehlers, J., Geren, P. \& Sachs, R. K. 1968, J. Math. Phys. 9, 1344. \\

\noindent
Ellis, G. F. R., Matravers, D. R. \& Treciokas, R. 1983, Ann. Phys. (N. Y.),
150, 455. \\

\noindent
Ellis, G. F. R., Treciokas, R., \& Matravers, D. R. 1983, Ann. Phys. (N. Y.),
150, 487. \\

\noindent
Gebbie, T., Ellis, G. F. R.,1998 astro-ph/9804316. \\

\noindent
Goodman, J. 1995, Phys. Rev. D, 52, 1821. \\

\noindent
Grishchuk, L. P., \& Zel'dovich, Ya. B. 1978, Sov. Astron. AJ, 22, 125. \\

\noindent
Hawking, S. W., \& Ellis, G. F. R. 1973, {\it The Large Scale Structure of
Space-Time}, Cambridge University Press, pp. 85-88. \\

\noindent
Kogut, A. et al. 1996, COBE preprint.

\noindent
Liddle, A. R., \& Lyth, D. H. 1993, Phys. Rep., 231, No. 1-2, 1 (pp. 57-59).\\

\noindent
Maartens, R., Ellis, G. F. R., \& Stoeger, W. R. 1995a, Phys. Rev. D, 51,
1525. \\

\noindent
Maartens, R., Ellis, G. F. R., \& Stoeger, W. R. 1995b, Phys. Rev. D, 51,
5942. \\

\noindent
Maartens, R., Ellis, G. F. R., \& Stoeger, W. R. 1996, Submitted to A\&A. \\

\noindent
Martinez-Gonzalez, E., \& Sanz, J. L. 1995, A\&A, 300, 346. \\

\noindent
Padmanabhan, T. 1993, {\it Structure Formation in the Universe}, Cambridge
University Press, pp 220ff. \\

\noindent
Partridge, R. B. 1995, {\it 3 K: The Cosmic Microwave Background Radiation},
Cambridge University Press, p. 189. \\

\noindent
Roos, M. 1994, {\it Introduction to Cosmology}, John Wiley and Sons, 
  p. 99 \\

\noindent
Smoot, G. F. et al. 1992, ApJ, 396, L1. \\

\noindent
Stoeger, W. R., Maartens, R., \& Ellis, G. F. R. 1995, ApJ, 443, 1. \\

\noindent
Wright, E. L. et al. 1994, ApJ, 436, 443. \\

\noindent
Zotov, N. V. and Stoeger, W. R. 1995, ApJ, 453, 574. \\

\end{document}